# Robust Beamforming for Security in MIMO Wiretap Channels with Imperfect CSI

Amitav Mukherjee, *Student Member, IEEE,* and A. Lee Swindlehurst, *Fellow, IEEE*

*Abstract*—In this paper, we investigate methods for reducing the likelihood that a message transmitted between two multi-antenna nodes is intercepted by an undetected eavesdropper. In particular, we focus on the judicious transmission of artificial interference to mask the desired signal at the time it is broadcast. Unlike previous work that assumes some prior knowledge of the eavesdropper's channel and focuses on maximizing secrecy capacity, we consider the case where no information regarding the eavesdropper is available, and we use signal-to-interference-plus-noise-ratio (SINR) as our performance metric. Specifically, we focus on the problem of maximizing the amount of power available to broadcast a jamming signal intended to hide the desired signal from a potential eavesdropper, while maintaining a prespecified SINR at the desired receiver. The jamming signal is designed to be orthogonal to the information signal when it reaches the desired receiver, assuming both the receiver and the eavesdropper employ optimal beamformers and possess exact channel state information (CSI). In practice, the assumption of perfect CSI at the transmitter is often difficult to justify. Therefore, we also study the resulting performance degradation due to the presence of imperfect CSI, and we present robust beamforming schemes that recover a large fraction of the performance in the perfect CSI case. Numerical simulations verify our analytical performance predictions, and illustrate the benefit of the robust beamforming schemes.

## I. INTRODUCTION

Due to their broadcast nature, wireless communications are inherently insecure. A passive eavesdropper within range of a wireless transmission obtains information about the transmitted signal without risk of detection. While encryption can be used to ensure confidentiality, its computational cost may be prohibitive and there are difficulties and vulnerabilities associated with key distribution and management [1]. Even when encryption is available, it is often still desirable to augment the security of the link and decrease the likelihood that its signals are detected or intercepted. As a result, there has recently been considerable interest in the use of physical layer mechanisms to increase the security of wireless communications systems.

Early work on the eavesdropper scenario, often referred to as the *wiretap* channel, focused on determining what conditions were necessary for secure communications in the presence of an eavesdropper [2]–[4]. In particular, this work led to the development of the notion of *secrecy capacity*, which quantifies the rate at which a transmitter can reliably send a secret message to the receiver, without the eavesdropper being able to decode it. Ultimately, for a wiretap channel without feedback, it was shown that a non-zero secrecy capacity can only be obtained if the eavesdropper's channel is of lower quality than that of the intended recipient. The work cited above assumed single antenna nodes; secrecy capacity for the multiple-antenna (MIMO) wiretap channel, where all nodes may possess multiple antennas, has been studied in [5]-[10].

A key consideration in the MIMO wiretap problem is what information is available about the eavesdropper. In principle, to compute the secrecy rate, one must know the eavesdropper's channel state information (CSI), or at least its distribution. Such information is unlikely to be available in many scenarios, especially those involving purely passive eavesdroppers. As a result, in this paper we take a different approach in which the transmitter minimizes the transmit power required to guarantee a certain Quality of Service (QoS) at the desired receiver, and uses the remaining resources to transmit an artificial interference signal that jams any eavesdroppers that are present [11], [12]. The use of artificial interference has been considered by a number of others even for the case where the eavesdropper's CSI is known, although such an approach is known to be suboptimal. For example, assuming that the transmitter has more antennas than the intended recipient so that the corresponding channel has a non-trivial nullspace, one of the approaches taken in [13] is to broadcast artificial interference in this nullspace. Such interference will have no impact on the receiver, but will in general degrade the eavesdropper's channel since its nullspace (if any) will be different. The high-SNR performance of this type of technique was shown to be nearly optimal in [6], and the optimal power distribution between data and interference has been examined in [14]. While [13] studied the case where only the distribution of the eavesdropper's channel was known, [6] focused on the situation where the transmitter has access to the eavesdropper's instantaneous CSI, and developed an algorithm to optimally exploit such information for the case where the intended recipient has a single antenna.

Another key consideration is the accuracy of the available CSI. The impact of imperfect CSI on the secrecy rate of the single-antenna wiretap channel has been investigated in [15], [16]. As we illustrate, techniques based on knowledge of the eavesdropper's channel in the multiple antenna case are very sensitive to even slight perturbations in the CSI. If unaccounted for, imprecise CSI for the primary channel also causes interference leakage to the desired recipient when artificial noise is used to jam the eavesdropper, resulting in significant degradation in the desired user's performance.

The authors are with the Dept. of Electrical Engineering & Computer Science, University of California, Irvine, CA 92697-2625, USA. (e-mail: {amukherj; swindle}@uci.edu)

A portion of this work was presented at the IEEE International Conference on Communications (ICC), Cape Town, South Africa, May 2010.

This work was supported by the U.S. Army Research Office under the Multi-University Research Initiative (MURI) grant W911NF-07-1-0318.

Consequently, we are interested in developing robust schemes that are insensitive to CSI errors. As such, we assume the transmitter uses beamforming rather than spatial multiplexing to communicate with the desired receiver. Beamforming is known to provide higher capacity than spatial multiplexing in many situations where the CSI at the transmitter is in the form of a mean and covariance (similar to the case considered here), even when the receiver has perfect CSI [17]. When the receiver CSI is also subject to errors, recent work has shown that beamforming is optimal even for small channel perturbations [18].

Since we focus on transmission of a single data stream using beamforming, and we let the received signal-to-interference-plus-noise-ratio (SINR) of the data stream at the desired receiver serve as our QoS metric. We design robust algorithms that minimize the transmit power required for the desired receiver to achieve the target QoS in the presence of CSI errors. This in turn maximizes the power available to transmit a jamming signal that disrupts the ability of the eavesdroppers to recover the desired signal. The robust algorithms rely on knowledge of the statistics of the CSI errors, and use a second-order perturbation analysis of the primary channel's singular value decomposition to account for the effects of the perturbation on the desired data stream. As a result, the algorithms provide the following benefits: (1) they minimize the effect of the jamming interference at the desired receiver when CSI errors are present, which means that (2) they require less transmit power to achieve the desired QoS, which in turn (3) maximizes the power available for degrading the channel of the eavesdroppers. Our simulations demonstrate that the resulting secrecy capacity is significantly improved compared with what would be obtained by a naive scheme that did not take CSI errors into account. We note that a similar approach can be taken to study the impact of imperfect CSI on schemes that make use of relays or neighboring users to jam eavesdroppers [19]-[23].

The paper is organized as follows. In the next section, the assumed mathematical model is presented, and the capabilities of the transmitter, receiver and eavesdropper are detailed. We also discuss the use of artificial interference, and examine the use of secrecy capacity and SINR as performance metrics. Fixed-QoS beamforming algorithms are described in Section III for the perfect CSI case, and the effects of imperfect CSI are analytically evaluated in Section IV. Robust beamforming methods that compensate for the degradation in SINR are then developed in Section V. The resulting SINR performance for a range of antenna configurations and CSI perturbations is studied via simulation in Section VI, and conclusions are drawn in Section VII.

## II. SYSTEM MODEL WITH PERFECT CSI

We assume a scenario with two cooperating nodes, Alice and Bob, and a passive eavesdropper, Eve. Each of the nodes may possess multiple antennas, the number of which we denote by $N_a$, $N_b$ and $N_e$, respectively. By the term "cooperating," we mean that Alice and Bob share information with each other about channel state information, desired link quality and coding/decoding strategies. Eve is non-cooperative in the sense that Alice and Bob are unaware of Eve's operating parameters, including her channel state information, number of antennas, etc. Alice is attempting to communicate a message to Bob in the presence of Eve, who is able to overhear Alice's transmissions. Eve need not be a single receiver with colocated antennas; our definition of "Eve" in this context could be multiple receivers in scattered locations who are able to coherently coordinate their received data. The signals received by Bob and Eve can be represented as follows:

$$\mathbf{y}_b = \mathbf{H}_{ba}\mathbf{x}_a + \mathbf{n}_b \qquad (1)$$
$$\mathbf{y}_e = \mathbf{H}_{ea}\mathbf{x}_a + \mathbf{n}_e, \qquad (2)$$

where $\mathbf{x}_a$ is the signal vector transmitted by Alice, $\mathbf{n}_b, \mathbf{n}_e$ are the naturally occurring noise and interference received by Bob and Eve, respectively, and $\mathbf{H}_{ba}, \mathbf{H}_{ea}$ are the corresponding $N_b \times N_a$ and $N_e \times N_a$ channel matrices. The channels $\mathbf{H}_{ba}, \mathbf{H}_{ea}$ are assumed to be deterministic quantities unrelated to each other, and no assumptions are made about their dimensions or structure.

The background noise is assumed to be spatially white, with possibly different power levels:

$$E\{\mathbf{n}_b\mathbf{n}_b^H\} = \sigma_b^2 \mathbf{I}$$
$$E\{\mathbf{n}_e\mathbf{n}_e^H\} = \sigma_e^2 \mathbf{I},$$

where $E\{\cdot\}$ denotes expectation, $(\cdot)^H$ the Hermitian transpose, and $\mathbf{I}$ is an identity matrix of appropriate dimension. The transmit power available for Alice is bounded by $P$:

$$E\{\mathbf{x}_a\mathbf{x}_a^H\} = \mathbf{Q}_a$$
$$\text{Tr}(\mathbf{Q}_a) \leq P,$$

where $\text{Tr}(\cdot)$ denotes the trace operator. Without loss of generality, we normalize $\mathbf{H}_{ba}$ so that its elements have unit-average gain (excess energy available from $\mathbf{H}_{ba}$ is assumed to be included in $P$):

$$\frac{\|\mathbf{H}_{ba}\|_F^2}{N_bN_a} = 1$$
$$\frac{\|\mathbf{H}_{ea}\|_F^2}{N_eN_a} = \gamma_{ea}^2 .$$

### A. Artificial Interference

Techniques that employ artificial interference devote a fraction of Alice's power to the transmission of a noise-like waveform, in an attempt to degrade the ability of Eve to intercept the signal destined for Bob. Since we are focusing on a beamforming scenario, Alice's signal is split into two components: one being a scalar data stream denoted as $z$ that contains the message for Bob, and one that contains the jamming signal, which we denote by the $N_a \times 1$ vector $\mathbf{z}'$. Bob therefore receives

$$\mathbf{y}_b = \mathbf{H}_{ba}\mathbf{t}z + \mathbf{H}_{ba}\mathbf{z}' + \mathbf{n}_b, \qquad (3)$$

where $\mathbf{t}$ is the $N_a \times 1$ transmit beamformer used for the information signal. Similarly, Eve sees

$$\mathbf{y}_e = \mathbf{H}_{ea}\mathbf{t}z + \mathbf{H}_{ea}\mathbf{z}' + \mathbf{n}_e. \qquad (4)$$

Assume $\mathbf{t}^H \mathbf{t} = 1$ and let $E\{|z|^2\} = \rho P$, where $0 < \rho \le 1$ is the fraction of the power devoted to the information signal, so that

$$
\begin{aligned}
E\{\mathbf{z}'\mathbf{z}'^H\} &= \mathbf{Q}'_z \\
\mathrm{Tr}(\mathbf{Q}'_z) &= (1-\rho)P .
\end{aligned}
$$

The QoS experienced by Bob and the probability of Eve intercepting the message intended for Bob will be determined by Alice's choice of the following parameters: the covariance matrix $\mathbf{Q}'_z$, the transmit beamformer $\mathbf{t}$, and the power allocation parameter $\rho$. The impact of these parameters on secrecy capacity and SINR are discussed in Sec. II-B.

It is important to note that the design of a complete transmission strategy for secrecy must also involve the construction of a "secrecy codebook" that is comprised of sub-codebooks for both the secret message and a randomization message intended to confuse the eavesdropper [24]. This is true even for situations where little or no information about the eavesdropper is present; in such cases, one can design the codebook using a set of worst-case assumptions about the eavesdropper. In a sense, the beamforming techniques discussed here represent a version of this idea in the spatial domain, where the secret and random messages are assigned to different spatial precoders (beamformers) with different transmit powers. An optimal design would presumably involve the joint construction of encoding schemes in both space and time, but such an effort is beyond the scope of this paper.

### B. Performance Metrics

Early work on the wiretap channel [2]–[4] led to the concept of secrecy capacity, which is defined to be the maximum rate at which Alice and Bob can communicate without allowing the eavesdropper to obtain any information about the transmitted message. In [7], it was shown that for the case where the background noise for Bob and Eve is of equal power (and no artificial interference is generated, $\mathbf{z}' = 0$), the secrecy capacity for the MIMO wiretap channel is given by

$$
\begin{aligned}
C_{sec} &= \max_{\mathbf{Q}_a \ge 0} I(\mathbf{X}_a; \mathbf{Y}_b) - I(\mathbf{X}_a; \mathbf{Y}_e) \quad (5) \\
&= \max_{\mathbf{Q}_a \ge 0} \log|\mathbf{I} + \mathbf{H}_{ba}\mathbf{Q}_a\mathbf{H}_{ba}^H| - \log|\mathbf{I} + \mathbf{H}_{ea}\mathbf{Q}_a\mathbf{H}_{ea}^H| ,
\end{aligned}
$$

where $I(\cdot; \cdot)$ represents mutual information, and where $\mathbf{Y}_b$, $\mathbf{Y}_e$ and $\mathbf{X}_a$ are the random variable counterparts to the specific realizations $\mathbf{y}_b, \mathbf{y}_e$ and $\mathbf{x}_a$, respectively. The secrecy-capacity-achieving choice for $\mathbf{Q}_a$ was derived in [7] for the case where the transmitter has knowledge of both $\mathbf{H}_{ba}$ and $\mathbf{H}_{ea}$, which were assumed to be fixed.

The use of secrecy capacity as the performance metric with artificial interference was studied in [13], where knowledge of only the distribution of $\mathbf{H}_{ea}$ was assumed and the expected value of (6) was maximized to obtain the ergodic secrecy capacity. The approach of [13] allowed for the transmission of multiple data streams to Bob, but restricted attention to the case where $N_a > N_b$, and forced Alice to choose a transmit covariance matrix according to the standard water-filling solution without regard to the possibility of an eavesdropper. The expected value of (6) was then maximized over $\rho$, where the expectation was taken over the distribution of eavesdropper channels, and it was assumed that $\sigma_e^2 = 0$. Note that, although this approach obviates the need for knowledge of Eve's instantaneous channel, optimization over $\rho$ still requires knowledge of the number of antennas Eve possesses and the strength of Eve's channel relative to Bob's (inherent in the assumption that the channel distribution is available).

Without any information about $\mathbf{H}_{ea}$, the above maximization problem is ill-posed, although (6) can still be used to quantify the secrecy rate of a given transmission scheme. In our work, we restrict attention to situations where Alice transmits only a single data stream to Bob since (1) we will focus on cases where the CSI is imperfectly known, and (2) we can develop methods that make beamforming robust to CSI errors. As a result, we choose to work directly with SINR rather than capacity. We will calculate the SINR assuming that both Bob and Eve use linear receive beamforming, recognizing the fact that both could use more sophisticated nonlinear techniques for decoding Alice's signal. The SINR achieved by linear beamforming will nonetheless provide an indication of the relative ability of Bob and Eve to determine the transmitted signal regardless of which decoding approach is used.

Let $\mathbf{w}_b$, $\mathbf{w}_e$ respectively denote the $N_b \times 1, N_e \times 1$ beamformers employed by Bob and Eve to determine $z$, so that

$$
\begin{aligned}
\hat{z}_b &= \mathbf{w}_b^H \mathbf{y}_b = \mathbf{w}_b^H \left( \mathbf{H}_{ba}\mathbf{t}z + \mathbf{H}_{ba}\mathbf{z}' + \mathbf{n}_b \right) \quad (6) \\
\hat{z}_e &= \mathbf{w}_e^H \mathbf{y}_e = \mathbf{w}_e^H \left( \mathbf{H}_{ea}\mathbf{t}z + \mathbf{H}_{ea}\mathbf{z}' + \mathbf{n}_e \right). \quad (7)
\end{aligned}
$$

The resulting SINR available for Bob and Eve to decode $z$ will be given by

$$
\mathrm{SINR}_b = \frac{\rho P |\mathbf{w}_b^H \mathbf{H}_{ba}\mathbf{t}|^2}{\mathbf{w}_b^H \left( \mathbf{H}_{ba}\mathbf{Q}'_z\mathbf{H}_{ba}^H + \sigma_b^2\mathbf{I} \right) \mathbf{w}_b} \quad (8)
$$

$$
\mathrm{SINR}_e = \frac{\rho P |\mathbf{w}_e^H \mathbf{H}_{ea}\mathbf{t}|^2}{\mathbf{w}_e^H \left( \mathbf{H}_{ea}\mathbf{Q}'_z\mathbf{H}_{ea}^H + \sigma_e^2\mathbf{I} \right) \mathbf{w}_e} . \quad (9)
$$

Intuitively, as long as $\mathrm{SINR}_b > \mathrm{SINR}_e$, there will exist modulation and coding schemes that allow Bob but not Eve to reliably decode $z$.

## III. FIXED-SINR BEAMFORMING WITH PERFECT CSI

In many applications, it is impractical to assume that any information about the eavesdropper's CSI is available. To increase communications security in such cases, we propose an approach that attempts to achieve the following two performance objectives: (1) maintain a certain guaranteed level of link quality (*e.g.*, SINR) for the intended receiver, and (2) maximize the power available for a jamming signal that makes the unintended reception of the signal more difficult. Obviously, the performance of such a scheme cannot be guaranteed; a fortuitous eavesdropper in the right location could end up with a better quality signal. Here the goal is to reduce the likelihood of such an event. Note that this approach does not imply that a low-power transmission from Alice to Bob will be more secure; reducing the power of the desired signal may allow one to better degrade Eve's channel, but it also reduces the requirements for Eve to decode the signal as well. To illustrate the proposed artificial interference concept, we assume here that the CSI is perfectly known by all parties, Alice, Bob and Eve. The case where Bob and Alice have imperfect or perturbed CSI is examined in Section IV.

### A. Unknown Eavesdropper CSI

The proposed approach can be generally outlined as follows, using SINR as the QoS metric:

1) Specify a target SINR for Bob.
2) Allocate the smallest possible fraction $\rho$ of the available transmit power to achieve the desired SINR (if possible) assuming Bob experiences no interference other than the background noise of power $\sigma_b^2$.
3) Allocate all of Alice's remaining power to a jamming signal that is uniformly distributed in space, subject to the constraint that when the interference is received by Bob, it lies in a subspace orthogonal to the desired signal.

Obviously, a given $\mathbf{H}_{ba}$ may not support the desired SINR with a total transmit power $P$; in such cases, the link is assumed to be in outage.

Let $S$ denote the target SINR for Bob. To minimize the fraction of the transmit power required to achieve $S$, Alice should choose $\mathbf{t}$ to be the right singular vector of $\mathbf{H}_{ba}$ with largest singular value, and Bob should choose $\mathbf{w}_b = \mathbf{H}_{ba}\mathbf{t}$ as his receive beamformer. Using this approach, we have

$$\rho = \frac{\sigma_b^2 S}{\mathbf{t}^H \mathbf{H}_{ba}^H \mathbf{H}_{ba} \mathbf{t} P} = \frac{\sigma_b^2 S}{\sigma_1^2 P}, \tag{10}$$

where $\sigma_1$ is the largest singular value of $\mathbf{H}_{ba}$. As long as $\rho < 1$, Alice has power available for generating artificial interference.

Since the CSI of the eavesdropper is unknown, the best option available to Alice is to uniformly spread the remaining transmit power along spatial dimensions that will produce no interference for Bob. In particular, we require that

$$\mathbf{H}_{ba}\mathbf{t} \perp \mathbf{H}_{ba}\mathbf{z}' \tag{11}$$

for all $\mathbf{z}'$. With $\mathbf{t}$ chosen as above, it is easy to see that $\mathbf{z}'$ must be chosen as a linear combination of the $N_a - 1$ right singular vectors of $\mathbf{H}_{ba}$ with smallest singular values, which we denote by $\mathbf{T}'$. Uniformly distributing the remaining transmit power over these vectors yields the following transmit covariance for the artificial interference:

$$\mathbf{Q}_z' = \frac{(1-\rho)P}{N_a - 1} \mathbf{T}'\mathbf{T}'^H. \tag{12}$$

As a consequence, the optimal (in the maximum SINR sense) receive beamformer for Bob is simply the maximal ratio combiner, $\mathbf{w}_b = \mathbf{H}_{ba}\mathbf{t}$, since Bob experiences only white noise. For Eve, the beamformer that maximizes SINR is given by

$$\mathbf{w}_e = \left(\mathbf{H}_{ea}\mathbf{Q}_z'\mathbf{H}_{ea}^H + \sigma_e^2\mathbf{I}\right)^{-1}\mathbf{H}_{ea}\mathbf{t}, \tag{13}$$

where $\mathbf{Q}_z'$ is given by (12). The use of an optimal beamformer here presumes that Eve is aware of $\mathbf{H}_{ea}\mathbf{t}$, as well as the spatial covariance matrix of the transmitted interference. With this

choice for $\mathbf{w}_e$, the SINR experienced by Eve can be expressed as

$$\text{SINR}_e = \rho P \mathbf{t}^H \mathbf{H}_{ea}^H \left(\mathbf{H}_{ea}\mathbf{Q}_z'\mathbf{H}_{ea}^H + \sigma_e^2\mathbf{I}\right)^{-1}\mathbf{H}_{ea}\mathbf{t}. \tag{14}$$

Since $\rho$ is proportional to $\sigma_b^2$, two observations are immediate for the case of low background noise ($\sigma_b^2, \sigma_e^2 \to 0$):

1) If $\mathbf{H}_{ea}\mathbf{Q}_z'\mathbf{H}_{ea}^H$ is full rank, which will generically be true if Alice has more antennas than Eve, then

$$\lim_{\sigma_b^2 \to 0} \text{SINR}_e = 0,$$

regardless of $\sigma_e^2$.

2) If $\mathbf{H}_{ea}\mathbf{Q}_z'\mathbf{H}_{ea}^H$ is rank deficient, for example if Eve has more antennas than Alice, then

$$\lim_{\sigma_e^2 \to 0} \left(\mathbf{H}_{ea}\mathbf{Q}_z'\mathbf{H}_{ea}^H + \sigma_e^2\mathbf{I}\right)^{-1} = \frac{1}{\sigma_e^2}\mathbf{R}\mathbf{R}^H,$$

where $\mathbf{R}$ is an orthonormal basis for the subspace orthogonal to $\mathbf{H}_{ea}\mathbf{Q}_z'^{1/2}$. In this case, if $\sigma_b^2 \to 0$ but $\sigma_b/\sigma_e \simeq O(1)$, then in general $\text{SINR}_e$ remains non-zero.

### B. Known Eavesdropper CSI

While our focus is on the case where Eve's CSI is unknown, it is useful to compare the performance of the artificial noise scheme with the optimal transmission strategy that takes knowledge of Eve's CSI into account. If perfect CSI of the eavesdropper's channel is available, then it is known that the use of artificial interference is suboptimal. The optimal approach to the problem posed in this paper is for Alice to transmit with full power using the beamformer that minimizes the eavesdropper's SINR given that the intended receiver's SINR is $S$:

$$\min_{\mathbf{t}} \text{SINR}_e \tag{15}$$
$$s.t. \text{SINR}_b = S.$$

It is straightforward to show that the solution to (15) is the generalized eigenvector $\mathbf{t}$ corresponding to the largest generalized eigenvalue $\lambda_{max}$ in the equation

$$\mathbf{H}_{ba}^H\mathbf{H}_{ba}\mathbf{t} = \lambda_{max}\mathbf{H}_{ea}^H\mathbf{H}_{ea}\mathbf{t}, \tag{16}$$

where $\mathbf{t}$ is scaled to ensure that $\text{SINR}_b = S$, provided that the transmit power $P$ is large enough. Clearly, if $N_e < N_a$, then $\mathbf{t}$ will lie in the nullspace of $\mathbf{H}_{ea}$ and $\text{SINR}_e = 0$. In such cases, it is preferable from a numerical point of view to calculate $\mathbf{t}$ as the generalized eigenvector with the smallest generalized eigenvalue in this equation:

$$\mathbf{H}_{ea}^H\mathbf{H}_{ea}\mathbf{t} = \lambda_{min}\mathbf{H}_{ba}^H\mathbf{H}_{ba}\mathbf{t}. \tag{17}$$

## IV. IMPACT OF IMPERFECT CSI

The assumption of perfect CSI at the transmitter is obviously impossible to achieve in practice. CSI uncertainty at Alice can be due to a number of different phenomena, including estimation error, quantized feedback, or channel mobility. CSI at the receiver is typically much more accurate, due to the receiver's ability to employ rapid channel tracking techniques based on, for example, decision direction. In this section, we examine the effect of inaccurate or mismatched

CSI between Alice and Bob using a second-order perturbation analysis of the singular value decomposition (SVD) of $\mathbf{H}_{ba}$, assuming that the channel error is described as a zero-mean random matrix with a given covariance. In the simulation section, we will demonstrate two important aspects of our analysis. First, we will show that the analysis accurately captures the effect of imperfect CSI even for relatively large channel errors, where the magnitude of the perturbation approaches that of the elements of the channel matrix itself. Second, our analysis will show that the previously proposed beamforming algorithms are very sensitive to imperfect CSI, and result in large degradations in SINR even when the channel perturbation is relatively small. This provides motivation for us to consider beamforming schemes that are robust to CSI errors, as developed in Section V.

For the analysis, we assume that $\mathbf{H}_{ba}$ is of full rank $F = \min(N_b, N_a)$, and we define the singular value decomposition of the unperturbed channel as follows:

$$
\begin{align}
\mathbf{H}_{ba} &= \mathbf{U}\mathbf{\Sigma}\mathbf{V}^H \tag{18} \\
&= [\mathbf{U}_s \; \mathbf{u}_F] \begin{bmatrix} \mathbf{\Sigma}_s & 0 \\ 0 & \sigma_F \end{bmatrix} [\mathbf{V}_s \; \mathbf{v}_F]^H \tag{19} \\
&= \mathbf{U}_s\mathbf{\Sigma}_s\mathbf{V}_s^H + \sigma_F\mathbf{u}_F\mathbf{v}_F^H , \tag{20}
\end{align}
$$

where $\mathbf{U}_s, \mathbf{V}_s$ contain respectively the first $F-1$ left and right singular vectors whose singular values are found in the diagonal matrix $\mathbf{\Sigma}_s$, and $\mathbf{u}_F, \mathbf{v}_F$ are respectively the left and right singular vectors corresponding to the smallest singular value $\sigma_F$. The partitioning of the SVD will be useful as we use the perturbation analysis of [25].

For purposes of our analysis, we assume that the CSI error is confined to Alice, who is assumed to have available the following perturbed channel estimate:

$$
\tilde{\mathbf{H}}_{ba} = \mathbf{H}_{ba} + \Delta\mathbf{H}_{ba} \tag{21}
$$

where $\Delta\mathbf{H}_{ba}$ is modeled as a zero-mean circularly-symmetric random matrix with covariance matrix given by

$$
\mathbf{C}_{\Delta\mathbf{H}_{ba}} = E\left\{ (\text{vec}(\Delta\mathbf{H}_{ba}))\,(\text{vec}(\Delta\mathbf{H}_{ba}))^H \right\} ,
$$

and $\text{vec}(\cdot)$ denotes the column stacking operator. The singular value decomposition of the perturbed channel can be written as

$$
\tilde{\mathbf{H}}_{ba} = \tilde{\mathbf{U}}_s\tilde{\mathbf{\Sigma}}_s\tilde{\mathbf{V}}_s^H + \tilde{\sigma}_F\tilde{\mathbf{u}}_F\tilde{\mathbf{v}}_F^H , \tag{22}
$$

where

$$
\begin{align}
\tilde{\mathbf{U}}_s &= \mathbf{U}_s + \Delta\mathbf{U}_s & \tilde{\mathbf{u}}_F &= \mathbf{u}_F + \Delta\mathbf{u}_F \\
\tilde{\mathbf{\Sigma}}_s &= \mathbf{\Sigma}_s + \Delta\mathbf{\Sigma}_s & \tilde{\sigma}_F &= \sigma_F + \Delta\sigma_F \\
\tilde{\mathbf{V}}_s &= \mathbf{V}_s + \Delta\mathbf{V}_s & \tilde{\mathbf{v}}_F &= \mathbf{v}_F + \Delta\mathbf{v}_F ,
\end{align} \tag{23}
$$

and quantities preceded by $\Delta$ are perturbations to those in (20). The analysis of [25] assumes either a fat or square matrix ($N_a \geq N_b$ in our case), so we perform our derivation for this case. A similar analysis holds when $N_b > N_a$, except that we would work with the transpose of the channel matrix, and we would focus on perturbations to the left rather than right singular vectors.

It will be convenient for our analysis to also define $\Delta\sigma_1$ and $\Delta\mathbf{v}_1$ as the perturbation to the largest singular value and the corresponding right singular vector $\mathbf{v}_1$, respectively. Furthermore, we define $\Delta\mathbf{T}'$ as the perturbation to the $N_a - 1$ right singular vectors of $\mathbf{H}_{ba}$ with smallest singular values. With $\Delta\sigma_1$ defined, the perturbed power allocation factor can be expressed as:

$$
\begin{align}
\tilde{\rho} &= \frac{\sigma_b^2 S}{\tilde{\sigma}_1^2 P} = \rho \frac{1}{\left(1 + \frac{2\sigma_1\Delta\sigma_1 + \Delta\sigma_1^2}{\sigma_1^2}\right)} \tag{24} \\
&\approx \rho\left(1 - \frac{2\Delta\sigma_1}{\sigma_1} - \frac{\Delta\sigma_1^2}{\sigma_1^2}\right), \tag{25}
\end{align}
$$

If Alice has an inaccurate estimate of the CSI and both Alice and Bob are unaware of the CSI mismatch, then the SINR for Bob is expected to be significantly degraded. There are three factors that contribute to this degradation:

1) Alice will incorrectly allocate power for data and artificial noise based on $\tilde{\rho} = (\sigma_b^2 S)/(\tilde{\sigma}_1^2 P)$.
2) Alice continues to use (12) to generate the interference signal, although with imperfect CSI the artificial noise covariance matrix becomes

$$
\tilde{\mathbf{Q}}_z' = \frac{(1-\tilde{\rho})P}{N_a - 1}(\mathbf{T}' + \Delta\mathbf{T}')(\mathbf{T}' + \Delta\mathbf{T}')^H . \tag{26}
$$

3) Alice will use $\mathbf{t} = \tilde{\mathbf{v}}_1 = \mathbf{v}_1 + \Delta\mathbf{v}_1$ as the transmit beamformer, whereas Bob continues to use $\mathbf{w}_b = \mathbf{H}_{ba}\mathbf{v}_1$ as his receive beamformer. Bob's beamformer will no longer cancel the artificial interference, causing a significant loss of SINR and the bulk of the resulting performance degradation.

This case of mismatched beamformers and erroneous power allocation due to imperfect CSI is referred to as the "*naive*" scheme.

In the presence of CSI errors, Bob's average SINR can be approximated as the ratio of the expected value of the received signal power to the expected value of the received noise and interference power. This approximation is valid to the order of the perturbation analysis assumed in [25], and its accuracy will be demonstrated later in our simulation results. Using this approximation, the average SINR achieved by Bob under the naive scheme can be expressed as

$$
\text{SINR}_b^{naive} = \frac{PE\left\{\tilde{\rho}|\mathbf{v}_1^H\mathbf{H}_{ba}^H\mathbf{H}_{ba}(\mathbf{v}_1 + \Delta\mathbf{v}_1)|^2\right\}}{E\left\{\mathbf{v}_1^H\mathbf{H}_{ba}^H\left(\mathbf{H}_{ba}\tilde{\mathbf{Q}}_z'\mathbf{H}_{ba}^H + \sigma_b^2\mathbf{I}\right)\mathbf{H}_{ba}\mathbf{v}_1\right\}}, \tag{27}
$$

where the remaining expectation is with respect to $\Delta\mathbf{H}_{ba}$. Based on the distribution of $\Delta\mathbf{H}_{ba}$, we can compute

$$
\begin{align}
E\left\{\mathbf{v}_1^H\mathbf{H}_{ba}^H\mathbf{H}_{ba}\tilde{\mathbf{Q}}_z'\mathbf{H}_{ba}^H\mathbf{H}_{ba}\mathbf{v}_1\right\} &= \sigma_1^4\tilde{\beta}E\left\{\mathbf{v}_1^H\tilde{\mathbf{T}}'\left(\tilde{\mathbf{T}}'\right)^H\mathbf{v}_1\right\} \tag{28} \\
&= \sigma_1^4\tilde{\beta}E\left\{\mathbf{v}_1^H\left(\mathbf{I} - \tilde{\mathbf{v}}_1\tilde{\mathbf{v}}_1^H\right)\mathbf{v}_1\right\} \\
&\approx -\sigma_1^4\beta E\left\{\mathbf{v}_1^H\Delta\mathbf{v}_1 + \Delta\mathbf{v}_1^H\mathbf{v}_1\right\},
\end{align}
$$

where $\tilde{\beta} = (1 - \tilde{\rho})P/(N_a - 1)$ and $\beta = (1 - \rho)P/(N_a - 1)$.

Let $\Upsilon = \frac{2\Delta\sigma_1}{\sigma_1} + \frac{\Delta\sigma_1^2}{\sigma_1^2}$. Using the familiar relations $\mathbf{H}_{ba}\mathbf{v}_1 = \sigma_1\mathbf{u}_1$ and $\mathbf{H}_{ba}^H\mathbf{u}_1 = \sigma_1\mathbf{v}_1$, and after dropping higher-order perturbation terms from the numerator and denominator, we obtain the following expression for SINR$_b^{naive}$:

$$\frac{\sigma_1^2 \rho P \left[1 + E\left\{\mathbf{v}_1^H \Delta \mathbf{v}_1\right\} + E\left\{\Delta \mathbf{v}_1^H \mathbf{v}_1\right\} - E\left\{\Upsilon\right\}\right]}{-\sigma_1^2 \beta \left[E\left\{\mathbf{v}_1^H \Delta \mathbf{v}_1\right\} + E\left\{\Delta \mathbf{v}_1^H \mathbf{v}_1\right\}\right] + \sigma_b^2}. \quad (29)$$

It is apparent that when perfect CSI is available at Alice (i.e., $\Delta \mathbf{v}_1 \to 0$ and $\Delta \sigma_1 \to 0$), (29) reduces to (10).

Next, we obtain the expected values of the perturbation terms $\mathbf{v}_1^H \Delta \mathbf{v}_1$, $\Delta \sigma_1$, and $\Delta \sigma_1^2$ in (29), the derivations of which are relegated to the Appendix. For convenience, let $\mathbf{C}_{ij} = E\left\{(\Delta \mathbf{H}_{ba})_{:,i} (\Delta \mathbf{H}_{ba})_{:,j}^H\right\}$ represent the covariance of columns $i$ and $j$ from $\Delta \mathbf{H}_{ba}$, i.e., $\mathbf{C}_{ij}$ is the $(i,j)$ block of $\mathbf{C}_{\Delta \mathbf{H}_{ba}}$. We also define the matrix $\mathbf{G}$ whose $(i,j)$ entry is given by $[\mathbf{G}]_{i,j} = \mathbf{v}_F^H \mathbf{C}_{ij} \mathbf{v}_F$. The expressions needed to evaluate Bob's SINR are given in (30)-(36):

$$\begin{aligned}
E\left\{[\mathbf{V}_s^H \Delta \mathbf{V}_s]\right\} &= -\frac{\sigma_F^2}{2} \mathbf{D} \mathbf{V}_s^H \mathbf{G} \mathbf{V}_s \mathbf{D} \quad (30)\\
&\quad - \frac{\sigma_F^2}{2} \mathbf{\Sigma}_s^{-1} \mathbf{D} \left(\left[(\sigma_F^2 + 1)\mathbf{I} + \mathbf{D}^{-1}\right] \times \right.\\
&\quad \left. \ldots \times \mathbf{U}_s^H \mathbf{G} \mathbf{U}_s \mathbf{D} + \mathbf{D}^{-1} \mathbf{U}_s^H \mathbf{G} \mathbf{U}_s\right) \mathbf{\Sigma}_s^{-1}\\
E\left\{\Delta \mathbf{\Sigma}_s\right\} &\approx (\sigma_F^2 \mathbf{U}_s^H \mathbf{G} \mathbf{U}_s \mathbf{D} + \mathbf{U}_s^H \mathbf{G} \mathbf{U}_s) \mathbf{\Sigma}_s^{-1} \quad (31)\\
&\quad - \mathbf{D} \left(\mathbf{\Sigma}_s \mathbf{V}_s^H \mathbf{G} \mathbf{V}_s \mathbf{\Sigma}_s + \ldots \right.\\
&\quad \left. \ldots + \sigma_F^2 \mathbf{U}_s^H \mathbf{G} \mathbf{U}_s\right) \mathbf{D} \mathbf{\Sigma}_s\\
&\quad + \mathbf{\Sigma}_s E\left[\mathbf{V}_s^H \Delta \mathbf{V}_s\right]\\
E\left\{\mathbf{v}_1^H \Delta \mathbf{v}_1\right\} &= E\left\{[\mathbf{V}_s^H \Delta \mathbf{V}_s]\right\}_{1,1} \quad (32)\\
\mathbf{D} &= \left(\mathbf{\Sigma}_s \mathbf{\Sigma}_s^H - \sigma_F^2 \mathbf{I}\right)^{-1} \quad (33)\\
E\left\{\Delta \sigma_1\right\} &= E\left\{[\Delta \mathbf{\Sigma}_s]_{1,1}\right\} \quad (34)\\
E\left\{\Delta \sigma_1^2\right\} &= \left[\mathbf{U}_s^H \mathbf{K} \mathbf{U}_s\right]_{1,1} \quad (35)\\
[\mathbf{K}]_{i,j} &= \text{Tr}\left(\mathbf{V}_s^H \mathbf{C}_{ij} \mathbf{V}_s\right). \quad (36)
\end{aligned}$$

Therefore, the naive SINR at Bob expressed in terms of the second-order statistics of $\Delta \mathbf{H}_{ba}$ is obtained by substituting the expected values in (32), (34), and (32) into (29). For the special case of i.i.d CSI errors where $\mathbf{C}_{\Delta H_{ba}} = \sigma_H^2 \mathbf{I}$, the expressions above simplify considerably since in this case $\mathbf{G} = \sigma_H^2 \mathbf{I}$.

Note that Alice's use of imperfect transmit beamformers does not implicitly impact the SINR available to Eve. As far as Eve is concerned, use of $\mathbf{v}_1 + \Delta \mathbf{v}_1$ rather than $\mathbf{v}_1$ as the transmit beamformer for the desired signal, and $\mathbf{T}' + \Delta \mathbf{T}'$ rather than $\mathbf{T}'$ as the interference precoder, has on average no effect on her performance since we assume that $\mathbf{H}_{ba}$ and $\mathbf{H}_{ea}$ are unrelated.

## V. ROBUST BEAMFORMING APPROACHES

While the instantaneous CSI perturbation cannot be determined, if Bob has information about the statistics of the perturbation, then he may take remedial measures to overcome at least some of the significant SINR degradation that occurs with the naive scheme. In particular, if Bob has knowledge of $\mathbf{C}_{\Delta \mathbf{H}_{ba}}$, then the spatial covariance of the artificial interference that impacts Bob can be calculated, and incorporated into the maximum SINR beamformer. In this section, we examine two such approaches for the case where Alice does not possess

CSI for Eve. The first case corresponds to a frequency-division duplex (FDD) scenario where Bob estimates the CSI, quantizes it, and sends this information to Alice via a feedback channel. In this case, Bob is aware of the CSI used by Alice for her transmission parameters. In the second case, which corresponds to a time-division duplex (TDD) scenario, Alice and Bob obtain individual channel estimates on their own, and neither is aware of the other's CSI. In both cases, we assume that (1) Alice's transmission allows Bob to obtain an exact estimate of the current CSI $\mathbf{H}_{ba}$ (the estimation error will be negligible compared with errors due to quantization and channel time variations), and that (2) Bob informs Alice of the power fraction $\rho$ needed to obtain his desired SINR.

### A. Robust Beamforming - FDD Case

When Alice has imperfect CSI for Bob and applies a mismatched transmit beamformer, the interference-plus-noise portion of Bob's received signal is, from (3),

$$\tilde{\mathbf{n}}_b = \mathbf{H}_{ba} \mathbf{z}' + \mathbf{n}_b,$$

with covariance

$$E\left\{\tilde{\mathbf{n}}_b \tilde{\mathbf{n}}_b^H\right\} = \mathbf{Q}_{int}. \quad (37)$$

In the FDD case, Bob is aware of the value of $\tilde{\mathbf{H}}_{ba}$ since this was information he computed and fed back to Alice. He can thus determine the exact value of $\mathbf{Q}_{int}$ as follows:

$$\mathbf{Q}_{int} = \tilde{\mathbf{H}}_{ba} \tilde{\mathbf{Q}}_z' \tilde{\mathbf{H}}_{ba}^H + \sigma_b^2 \mathbf{I}, \quad (38)$$

as well as the exact beamformer $\tilde{\mathbf{t}} = \tilde{\mathbf{v}}_1$ that Alice uses for the information-bearing signal. He is then in turn able to calculate the optimal receive beamformer that maximizes SINR:

$$\mathbf{w}_{opt} = \mathbf{Q}_{int}^{-1} \mathbf{H}_{ba} \tilde{\mathbf{t}}. \quad (39)$$

The resulting SINR at Bob is given by

$$S = \rho P \tilde{\mathbf{t}}^H \mathbf{H}_{ba}^H \mathbf{Q}_{int}^{-1} \mathbf{H}_{ba} \tilde{\mathbf{t}}. \quad (40)$$

### B. Robust Beamforming - TDD Case

In the TDD case, Bob is unaware of the exact values of $\tilde{\mathbf{Q}}_z'$ and $\tilde{\mathbf{v}}_1$ that Alice uses. However, assuming Bob knows the statistics of the CSI error, in particular $\mathbf{C}_{\Delta \mathbf{H}_{ba}}$, he can compute expected values for these quantities and use these as estimates to determine his receive beamformer. Using the second-order perturbation analysis of the previous section, the expected interference-plus-noise covariance matrix $\hat{\mathbf{Q}}_{int}$ can be computed as

$$\begin{aligned}
\hat{\mathbf{Q}}_{int} &= E\left\{\mathbf{H}_{ba} \mathbf{z}' \mathbf{z}'^H \mathbf{H}_{ba}^H + \mathbf{n}_b \mathbf{n}_b^H\right\} \quad (41)\\
&= \tilde{\beta} \left(\mathbf{H}_{ba} \mathbf{H}_{ba}^H - \sigma_1^2 \mathbf{u}_1 \mathbf{u}_1^H\right) - \beta \sigma_1^2 \mathbf{u}_1 E\left\{\Delta \mathbf{v}_1\right\} \mathbf{H}_{ba}^H\\
&\quad - \beta \sigma_1 \mathbf{H}_{ba} E\left\{\Delta \mathbf{v}_1\right\} \mathbf{u}_1^H + \sigma_b^2 \mathbf{I}.
\end{aligned}$$

Furthermore, Alice's transmit beamformer can be estimated as

$$\hat{\mathbf{t}} = E\left(\tilde{\mathbf{v}}_1\right) = \mathbf{v}_1 + E\left(\Delta \mathbf{v}_1\right). \quad (42)$$

Both of the above quantities require knowledge of $\Delta \mathbf{v}_1$. In the Appendix, we show that

$$E\left\{\Delta \mathbf{v}_1\right\} = E\left\{[\Delta \mathbf{V}_s]_{:,1}\right\}, \quad (43)$$

where

$$E\left\{\Delta\mathbf{V}_s\right\} = \mathbf{v}_F E\left\{\bar{\mathbf{P}}_1\right\} + \mathbf{V}_s E\left\{\bar{\mathbf{P}}_2\right\} \tag{44}$$

$$\begin{aligned}E\left\{\bar{\mathbf{P}}_1\right\} &= (1+\sigma_F^2)\mathbf{v}_F^H\mathbf{G}\mathbf{V}_s\mathbf{D}^H + \sigma_F^2\mathbf{v}_F^H\mathbf{G}''\mathbf{V}_s\mathbf{D}^H \\ &\quad - \sigma_F\mathbf{u}_F^H\mathbf{G}'\mathbf{U}_s\left(\mathbf{I}+\sigma_F^2\mathbf{D}^H\right)\boldsymbol{\Sigma}_s^{-1} \\ &\quad + \sigma_F\mathbf{u}_F^H\mathbf{G}\mathbf{U}_s\mathbf{D}^H\left(\sigma_F^2\mathbf{D}^H+\mathbf{I}\right)\boldsymbol{\Sigma}_s^{-1} \end{aligned} \tag{45}$$

$$[\mathbf{G}]_{i,j} = \mathbf{v}_F^H\mathbf{C}_{ij}\mathbf{v}_F \tag{46}$$

$$[\mathbf{G}']_{i,j} = \mathrm{Tr}\left(\mathbf{V}_s\mathbf{D}^H\mathbf{V}_s^H\mathbf{C}_{ij}\right) \tag{47}$$

$$[\mathbf{G}'']_{i,j} = \mathrm{Tr}\left(\mathbf{U}_s\mathbf{D}^H\mathbf{U}_s^H\mathbf{C}_{ij}\right)\ , \tag{48}$$

and where the expected value of $\bar{\mathbf{P}}_2 = \mathbf{V}_s^H\Delta\mathbf{V}_s$ is given in (30). The interference-plus-noise covariance matrix is obtained by substituting (43) into (42). Bob's receive beamformer is calculated as

$$\hat{\mathbf{w}}_{opt} = \hat{\mathbf{Q}}_{int}^{-1}\mathbf{H}_{ba}\hat{\mathbf{t}}\ . \tag{49}$$

Since $\hat{\mathbf{Q}}_{int} \neq \mathbf{Q}_{int}$, the resulting SINR for Bob must be determined as follows:

$$\mathrm{SINR}_b = \frac{\rho P\left|\hat{\mathbf{t}}^H\mathbf{H}_{ba}^H\hat{\mathbf{Q}}_{int}^{-1}\mathbf{H}_{ba}\hat{\mathbf{t}}\right|^2}{\hat{\mathbf{t}}^H\mathbf{H}_{ba}^H\hat{\mathbf{Q}}_{int}^{-1}\mathbf{Q}_{int}\hat{\mathbf{Q}}_{int}^{-1}\mathbf{H}_{ba}\hat{\mathbf{t}}}\ . \tag{50}$$

## VI. SIMULATION RESULTS

We present some examples that show the SINR and secrecy capacity performance of Bob and Eve for various array sizes, target performance levels, and array perturbations. In all simulations, the channel matrices were assumed to be composed of independent, zero-mean Gaussian random variables with unit variance ($\gamma_{ba}^2 = 1$). The channel perturbation covariance matrix is assumed to be $\mathbf{C}_{\Delta H_{ba}} = \sigma_H^2\mathbf{I}$ which corresponds to the case where the CSI errors are independent and identically distributed. In the simulation plots, $\sigma_H$ is specified in dB according to $20\log_{10}\sigma_H$. For example, a value of $\sigma_H = -20$dB corresponds to $\sigma_H = 0.1$, indicating channel perturbations on the order of 10% of the channel coefficients themselves. All displayed results are calculated based on an average of 3000 independent trials. The background noise power was assumed to be the same for both Bob and Eve: $\sigma_b^2 = \sigma_e^2 = 1$, and in all cases the available transmit power was assumed to be $P = 100$, or 20dB. In situations where the desired SINR for Bob cannot be achieved with the given $P$, rather than indicate an outage, we simply assign all power to Bob and zero to artificial interference and average the resulting SINR with the others.

### A. Effects of Eavesdropper CSI

Figure 1 illustrates the performance of the algorithms when $S = 20$dB and $N_e \in [1,20]$. The number of antennas for Alice and Bob are assumed to be equal, and results are shown for $N_a = N_b = 4,8$. The desired SINR for Bob was set to 20dB, and the available transmit power was sufficient in this simulation for the target to be met in all 3000 trials. Three curves are included for Eve, showing the performance of the algorithms for different assumptions about the eavesdropper's CSI (ECSI): (1) when it is unknown, in which case the artificial noise approach of Section III-A is used, (2) when it is perfectly known, in which case the generalized eigenvector approach of Section III-B is used, and (3) when it is imperfectly known, where again the approach of Section III-B is used, but the ECSI perturbation is unaccounted for. The perturbed ECSI was generated by the following equation, assuming $\gamma = 0.05$ (which corresponds to a perturbation of about -13dB):

$$\tilde{\mathbf{H}}_{ea} = \sqrt{1-\gamma}\mathbf{H}_{ea} + \sqrt{\gamma}\mathbf{W}_{ea}\ , \tag{51}$$

where $\mathbf{H}_{ea}$ and $\mathbf{W}_{ea}$ are independent, zero-mean Gaussian with unit-variance elements, and hence so is $\tilde{\mathbf{H}}_{ea}$. In the simulations, the actual channel is $\mathbf{H}_{ea}$, but Alice assumes it is $\tilde{\mathbf{H}}_{ea}$. The assumption of perfect ECSI provides a significant benefit when $N_e < \{N_a, N_b\}$; in fact, the eavesdropper's SINR can theoretically be driven to zero. The gain when $N_e \geq \{N_a, N_b\}$ is not as large, particularly for $N_a = N_b = 4$, when it is less than 2dB. Much of the benefit of ECSI is lost however if it is imprecisely known; even for this case when the perturbation is relatively small, we see that for small $N_e$ it is often better to ignore the ECSI than to use a perturbed version of it.

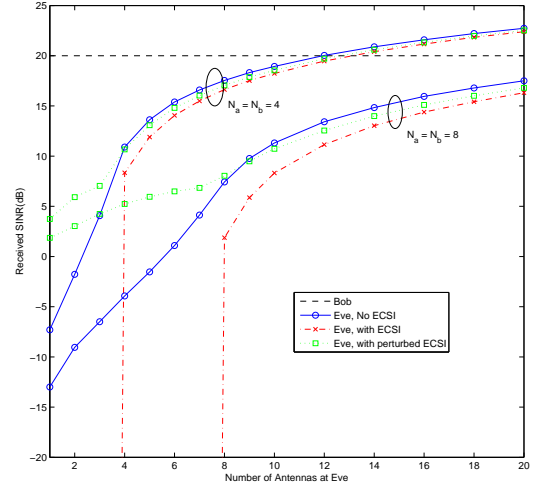

Fig. 1. SINR versus number of antennas for Eve.

### B. SINR Degradation Analysis

In Figure 2, we compare the SINR expressions for the naive case based on second-order perturbation theory derived in Section IV with measured SINR values from simulations for a range of channel perturbation powers. The set of channel matrices have dimensions of either $N_a = N_b = N_e = 2$ or $N_a = N_b = N_e = 5$, and the desired SINR for Bob is set to $S = 20$dB. For both antenna configurations, the second-order approximations appear to be accurate up to about $\sigma_H = -10$dB, which corresponds to $\sigma_H = 0.32$. This is a relatively large perturbation for channels with unit-variance elements. We see that inaccurate CSI substantially impacts Bob's SINR, even for relatively small values of $\sigma_H$. For example, when $N_a = 10$, Bob loses 6dB of SINR for the relatively small value $\sigma_H = 0.1$.

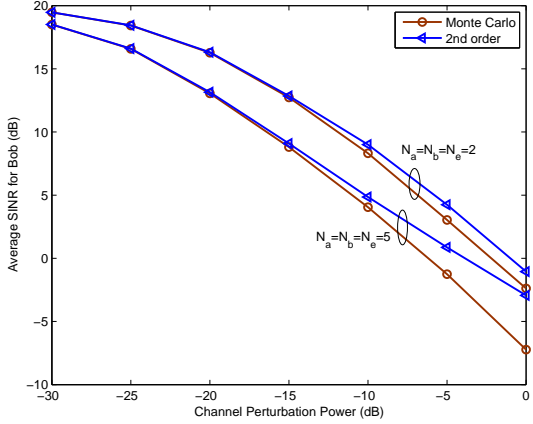

Fig. 2.   A comparison of the $2^{nd}$-order naive SINR approximations with Monte Carlo SINR results for $N_a = N_b = N_e = 2$ and $N_a = N_e = N_b = 5$.

### C. Robust Beamforming Results

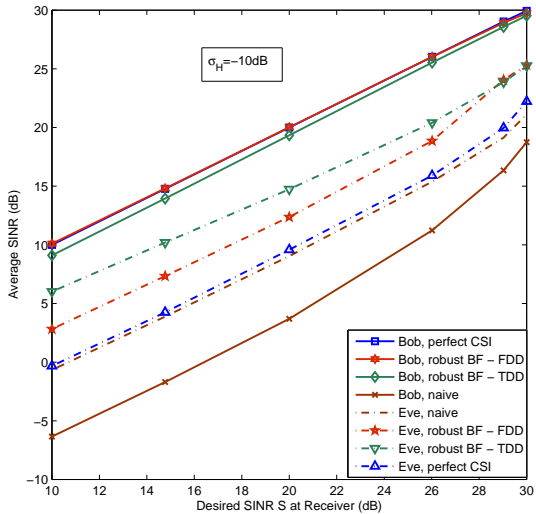

Fig. 3.   Measured SINR values versus desired SINR for Bob and Eve with perfect and imperfect CSI at Alice for $N_a = N_b = N_e = 5$ antennas, $\sigma_H = -10$dB.

Figure 3 shows the SINR for Bob and Eve as a function of $S$ for various approaches, including the robust beamforming schemes presented earlier. The channel perturbation power is fixed at $\sigma_H = -10$dB, and we assume $N_a = N_b = N_e = 5$. It is evident that the naive schemes incur a significant SINR penalty for relatively small channel perturbations, with the achieved SINR at the intended receiver being 15-17dB below the target SINR and 6-7dB worse than the SINR for Eve. Note however that the robust receive beamforming schemes are able to restore Bob's SINR performance at or near the desired value. Obviously, the presence of uncancelled artificial interference due to imperfect CSI requires Alice to use additional power for the desired signal, thus reducing the amount of noise available to jam the eavesdropper. This is why Eve's SINR increases with the robust beamforming methods. As expected, Eve's performance is best degraded in the FDD case where Bob has exact knowledge of Alice's transmission scheme[1]. Note also that Eve's SINR increases slightly for high values of $S$. This is due to the fact that as $S$ increases, there will be an increasing number of cases where no power is available for jamming. This also inadvertently helps Bob in the naive case, since the lack of jamming eliminates interference for the desired signal.

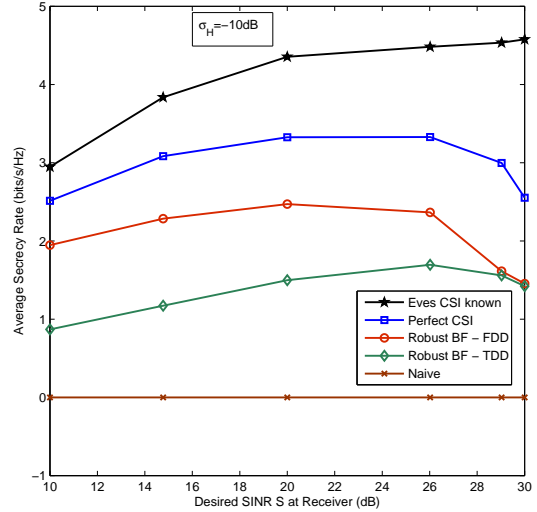

Fig. 4.   Secrecy capacity versus desired SINR for Bob with perfect and imperfect CSI at Alice for $N_a = N_b = N_e = 5$ antennas, $\sigma_H = -10$dB.

Figure 4 plots the secrecy capacity that results for the case considered here, for various CSI assumptions. The case where Eve's CSI is perfectly known is shown for reference, and obviously for this case the best secrecy capacity is obtained. As expected, the benefit of knowing the eavesdropper's CSI is largest when Bob demands a high QoS, and minimal for low values of $S$ where more resources are available for jamming. The robust beamforming strategies provide non-zero secrecy capacity for all values of $S$, and recover a reasonable fraction of the performance available in the perfect CSI case. However, in the naive case, the secrecy capacity is reduced to zero since Eve's SINR is always larger than Bob's. This assumes of course that Bob does nothing to counteract the interference, while Eve uses an optimal beamformer that requires exact knowledge of the interference covariance.

The effect of the magnitude of the channel perturbation on SINR performance is illustrated in Figure 5 for the case studied in the previous figures, assuming $S = 20$dB. Robust beamforming in the FDD case realizes little performance loss for values of $\sigma_H$ up to -15dB, while the threshold for degradation in the TDD case is somewhat lower. Recall that Figure 4 showed a positive secrecy capacity for the TDD case at $\sigma_H = -10$dB, even though in Figure 5 both Bob and Eve appear to have approximately the same average SINR. This is because the secrecy capacity must be non-negative; a positive result is obtained when Bob's SINR exceeds Eve's, but the capacity is assumed to be zero otherwise.

---

[1]This does not imply that FDD systems are better than TDD systems for this application; one may expect that in practice the value for $\sigma_H$ will be somewhat larger in the FDD case due to quantization and the added delay required for feedback.

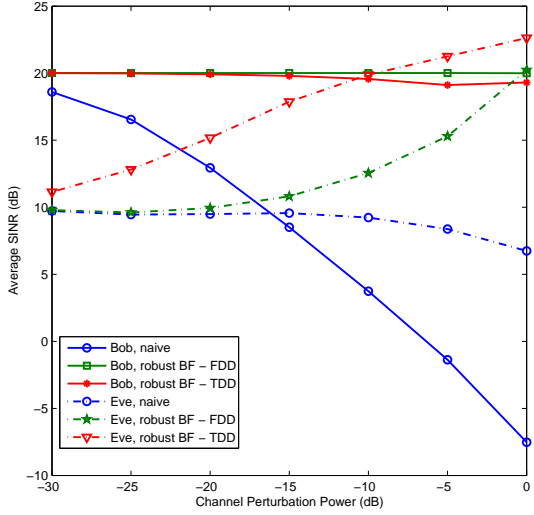

Fig. 5. Average SINR for Bob and Eve as a function of $\sigma_H$ for $N_a = N_b = N_e = 5$ antennas and $S = 20$dB.

## VII. CONCLUSIONS

We have presented beamforming-based approaches for improving the security of the wireless communications between two multi-antenna nodes. The algorithms allocate transmit power in order to achieve a target SINR for a desired user, and then broadcast the remaining available power as artificial noise in order to disrupt the interception of the signal by a passive eavesdropper. The proposed approaches rely heavily on the availability of accurate CSI, and their performance can be quite sensitive to imprecise channel estimates. As a result, we conducted a detailed second-order perturbation analysis in order to precisely quantify the effects of inaccurate CSI. Simulations were used to demonstrate the validity of the analysis, and to illustrate the sensitivity of algorithms that depend on precise CSI. To reduce the impact of the CSI errors, we proposed two robust beamforming schemes that are able to recover a large fraction of the SINR lost due to the channel estimation errors. These techniques were shown to perform very well for moderate CSI errors, but ultimately a large enough channel mismatch can eliminate the secrecy advantage of using artificial noise.

## APPENDIX

Define $\mathbf{D} \triangleq \left(\boldsymbol{\Sigma}_s \boldsymbol{\Sigma}_s^H - \sigma_F^2 \mathbf{I}\right)^{-1}$, as well as the following matrices:

$$\mathbf{E}_{ss} \triangleq \mathbf{U}_s^H \Delta \mathbf{H}_{ba} \mathbf{V}_s \qquad (52)$$
$$\mathbf{E}_{sn} \triangleq \mathbf{U}_s^H \Delta \mathbf{H}_{ba} \mathbf{v}_F \qquad (53)$$
$$\mathbf{E}_{ns} \triangleq \mathbf{u}_F^H \Delta \mathbf{H}_{ba} \mathbf{V}_s \qquad (54)$$
$$\mathbf{E}_{nn} \triangleq \mathbf{u}_F^H \Delta \mathbf{H}_{ba} \mathbf{v}_F \ . \qquad (55)$$

Using the results of [25], the perturbation in $\mathbf{V}_s$ can be approximated up to second order in $\Delta \mathbf{H}_{ba}$ as

$$\Delta \mathbf{V}_s = \mathbf{v}_F \bar{\mathbf{P}}_1 + \mathbf{V}_s \bar{\mathbf{P}}_2 \qquad (56)$$

where $\bar{\mathbf{P}}_1 \approx -\bar{\mathbf{Q}}_1^H$ and $\bar{\mathbf{P}}_2 \approx -\frac{1}{2}\bar{\mathbf{F}}\bar{\mathbf{F}}^H$, and

$$\bar{\mathbf{F}} = -\sigma_F \mathbf{D} \mathbf{E}_{ns}^H - \sigma_F^2 \boldsymbol{\Sigma}_s^{-1} \mathbf{D} \mathbf{E}_{sn} - \boldsymbol{\Sigma}_s^{-1} \mathbf{E}_{sn}, \qquad (57)$$

$$\begin{aligned}
\bar{\mathbf{Q}}_1 \approx\ & \mathbf{D}\left(\mathbf{E}_{ss}\mathbf{D}\boldsymbol{\Sigma}_s\mathbf{E}_{ns}^H - \sigma_F \mathbf{D}\mathbf{E}_{ns}^H\mathbf{E}_{nn}^H\right)\boldsymbol{\Sigma}_n \\
& - \mathbf{D}\mathbf{E}_{ns}^H\mathbf{E}_{nn} + \sigma_F^2 \mathbf{D}\left(\mathbf{E}_{ss}^H\mathbf{D}\mathbf{E}_{sn} - \mathbf{D}\mathbf{E}_{ns}^H\mathbf{E}_{nn}\right) \\
& + \sigma_F^2 \boldsymbol{\Sigma}_s^{-1}\mathbf{D}\left(\mathbf{E}_{ss}\boldsymbol{\Sigma}_s^{-1}\mathbf{E}_{sn} - \mathbf{D}\mathbf{E}_{sn}\boldsymbol{\Sigma}_n^H\mathbf{E}_{nn} + \ldots\right. \\
& \ldots + \left.\sigma_F^2\mathbf{E}_{ss}\boldsymbol{\Sigma}_s^{-1}\mathbf{D}\mathbf{E}_{sn}\right) \\
& + \boldsymbol{\Sigma}_s^{-1}\mathbf{E}_{ss}\boldsymbol{\Sigma}_s^{-1}\left(\mathbf{E}_{sn} + \sigma_F^2\mathbf{D}\mathbf{E}_{sn}\right) - \sigma_F\boldsymbol{\Sigma}_s^{-1}\mathbf{D}\mathbf{E}_{sn}\mathbf{E}_{nn} \\
& + \sigma_F\boldsymbol{\Sigma}_s^{-1}\mathbf{D}\left(\sigma_F^2\mathbf{E}_{ss}\mathbf{D}\mathbf{E}_{ns}^H - \sigma_F^2\mathbf{D}\mathbf{E}_{sn}\mathbf{E}_{nn}^H - \mathbf{E}_{sn}\mathbf{E}_{nn}^H\right) \\
& + \sigma_F\boldsymbol{\Sigma}_s^{-1}\mathbf{E}_{ss}\mathbf{D}\mathbf{E}_{ns}^H + \bar{\mathbf{F}}.
\end{aligned} \qquad (58)$$

Exploiting the circular symmetry of $\Delta \mathbf{H}_{ba}$ in (59) leads to

$$\begin{aligned}
E\{\bar{\mathbf{P}}_1\} =\ & (1+\sigma_F^2)E\{\mathbf{E}_{nn}^H\mathbf{E}_{ns}\}\mathbf{D}^H \\
& + \sigma_F^2 E\{\mathbf{E}_{sn}^H\mathbf{D}^H\mathbf{E}_{ss}\}\mathbf{D}^H \\
& - \sigma_F E\{\mathbf{E}_{ns}\mathbf{D}^H\mathbf{E}_{ss}^H\}\left(\mathbf{I} + \sigma_F^2\mathbf{D}^H\right)\boldsymbol{\Sigma}_s^{-1} \\
& + \sigma_F E\{\mathbf{E}_{nn}\mathbf{E}_{sn}^H\}\mathbf{D}^H\left(\sigma_F^2\mathbf{D}^H + \mathbf{I}\right)\boldsymbol{\Sigma}_s^{-1}.
\end{aligned} \qquad (59)$$

Next, recall that $\mathbf{V}_s \perp \mathbf{v}_F$ and $\mathbf{V}_s^H\mathbf{V}_s = \mathbf{I}$, so that $\mathbf{V}_s^H\Delta\mathbf{V}_s = \bar{\mathbf{P}}_2$. After some manipulations based on the circular symmetry of $\Delta\mathbf{H}_{ba}$, we obtain

$$\begin{aligned}
E\left[\mathbf{V}_s^H\Delta\mathbf{V}_s\right] =\ & -\frac{\sigma_F^2}{2}\boldsymbol{\Sigma}_s^{-1}\left((\sigma_F^2+1)\mathbf{D}E\{\mathbf{E}_{sn}\mathbf{E}_{sn}^H\}\mathbf{D}^H + \ldots\right. \\
& \ldots + \left.E\{\mathbf{E}_{sn}\mathbf{E}_{sn}^H\}\mathbf{D}^H + E\{\mathbf{E}_{sn}\mathbf{E}_{sn}^H\}\right)\boldsymbol{\Sigma}_s^{-1} \\
& - \frac{\sigma_F^2}{2}\mathbf{D}E\{\mathbf{E}_{ns}^H\mathbf{E}_{ns}\}\mathbf{D}^H.
\end{aligned} \qquad (60)$$

The perturbation to the singular values $\boldsymbol{\Sigma}_s$ can be approximated as

$$\begin{aligned}
E\{\Delta\boldsymbol{\Sigma}_s\} \approx\ & \left(\sigma_F^2 E\{\mathbf{E}_{sn}\mathbf{E}_{sn}^H\}\mathbf{D}^H + E\{\mathbf{E}_{sn}\mathbf{E}_{sn}^H\}\right)\boldsymbol{\Sigma}_s^{-1} \\
& + E\{\boldsymbol{\Sigma}_s\bar{\mathbf{P}}_2 - \mathbf{P}_2\boldsymbol{\Sigma}_s\},
\end{aligned} \qquad (61)$$

where $\mathbf{P}_2 \approx -\frac{1}{2}\mathbf{F}\mathbf{F}^H$ is a component of the perturbation in $\Delta\mathbf{U}_s$, and

$$\mathbf{F} = -\mathbf{D}\left(\boldsymbol{\Sigma}_s\mathbf{E}_{ns}^H + \sigma_F\mathbf{E}_{sn}\right). \qquad (62)$$

From the expression for $\mathbf{P}_2$:

$$\begin{aligned}
E\{\mathbf{P}_2\boldsymbol{\Sigma}_s\} =\ & \mathbf{D}\left(\boldsymbol{\Sigma}_s E\{\mathbf{E}_{ns}^H\mathbf{E}_{ns}\}\boldsymbol{\Sigma}_s \right. \\
& + \sigma_F^2 E\{\mathbf{E}_{sn}\mathbf{E}_{sn}^H\}\left.\right)\mathbf{D}^H\boldsymbol{\Sigma}_s.
\end{aligned} \qquad (63)$$

It remains to express (59) and (61) in terms of the second-order statistics of $\Delta\mathbf{H}_{ba}$. Let $\mathbf{C}_{ij} = E\left\{(\Delta\mathbf{H}_{ba})_{:,i}(\Delta\mathbf{H}_{ba})_{:,j}^H\right\}$ represent the covariance of the $i$th and $j$th columns of $\Delta\mathbf{H}_{ba}$. It is straightforward to

show that

$$E\left\{\mathbf{E}_{sn}\mathbf{E}_{sn}^H\right\} = \mathbf{U}_s^H E\left[\Delta\mathbf{H}_{ba}\mathbf{v}_F\mathbf{v}_F^H\Delta\mathbf{H}_{ba}^H\right]\mathbf{U}_s \tag{64}$$
$$= \mathbf{U}_s^H\mathbf{G}\mathbf{U}_s \tag{64}$$
$$E\left\{\mathbf{E}_{ns}^H\mathbf{E}_{ns}\right\} = \mathbf{V}_s^H\mathbf{G}''\mathbf{V}_s \tag{65}$$
$$E\left\{\mathbf{E}_{nn}\mathbf{E}_{sn}^H\right\} = \mathbf{u}_F^H\mathbf{G}\mathbf{U}_s \tag{66}$$
$$E\left\{\mathbf{E}_{nn}^H\mathbf{E}_{ns}\right\} = \mathbf{v}_F^H\mathbf{G}''\mathbf{V}_s \tag{67}$$
$$E\left\{\mathbf{E}_{ns}\mathbf{D}^H\mathbf{E}_{ss}^H\right\} = \mathbf{u}_F^H\mathbf{G}'\mathbf{U}_s \tag{68}$$
$$E\left\{\mathbf{E}_{sn}^H\mathbf{D}^H\mathbf{E}_{ss}\right\} = \mathbf{v}_F^H\mathbf{G}''\mathbf{V}_s\ , \tag{69}$$

where the $(i,j)$ entry of $\mathbf{G}$ is $[\mathbf{G}]_{i,j}=\mathbf{v}_F^H\mathbf{C}_{ij}\mathbf{v}_F$, $[\mathbf{G}']_{i,j}=\mathrm{Tr}\left(\mathbf{V}_s\mathbf{D}^H\mathbf{V}_s^H\mathbf{C}_{ij}\right)$, and $[\mathbf{G}'']_{i,j}=\mathrm{Tr}\left(\mathbf{U}_s\mathbf{D}^H\mathbf{U}_s^H\mathbf{C}_{ij}\right)$.

The required expected values in (32), (34), (32) and (43) immediately follow from the results derived above.